# Lattice dynamics in chiral tellurium by linear and circularly polarized Raman spectroscopy: crystal orientation and handedness

Davide Spirito,[1] Sergio Marras [2] and Beatriz Martín-García [3,4*]

[1]IHP–Leibniz-Institut für innovative Mikroelektronik, Im Technologiepark 25, 15236 Frankfurt (Oder), Germany.
[2]Istituto Italiano di Tecnologia - Materials Characterization Facility, Genova 16163, Italy.
[3]CIC nanoGUNE BRTA, 20018 Donostia-San Sebastián, Basque Country, Spain
[4]IKERBASQUE, Basque Foundation for Science, 48009 Bilbao, Spain.
Email:  b.martingarcia@nanogune.eu

## Abstract

Trigonal tellurium (Te) has attracted researchers' attention due to its transport and optical properties, which include electrical magneto-chiral anisotropy, spin polarization and bulk photovoltaic effect. It is the anisotropic and chiral crystal structure of Te that drive these properties, so the determination of its crystallographic orientation and handedness is key to their study. Here we explore the structural dynamics of Te bulk crystals by angle-dependent linearly polarized Raman spectroscopy and symmetry rules in three different crystallographic orientations. The angle-dependent intensity of the modes allows us to determine the arrangement of the helical chains and distinguish between crystallographic planes parallel and perpendicular to the chain axis. Furthermore, under different configurations of circularly polarized Raman measurements and crystal orientations, we observe the shift of two phonon modes only in the (0 0 1) plane. The shift is positive or negative depending on the handedness of the crystals, which we determine univocally by chemical etching. Our analysis of three different crystal faces of Te highlights the importance of selecting the proper orientation and crystallographic plane when investigating the transport and optical properties of this material. These results offer insight into the crystal structure and symmetry in other anisotropic and chiral materials, and open new paths to select a suitable crystal orientation when fabricating devices.

## Introduction

Materials with a chiral crystal structure have awakened the interest of researchers in the fields of physics, chemistry, and biology. These materials present an asymmetric structure, meaning they cannot be superimposed on their mirror image, giving rise to a pair of enantiomers with opposite handedness. Many examples can be found in organic molecules, where chirality is crucial for their functionality.[1–8] By contrast, few inorganic materials display intrinsically chiral crystal structures, however, those that do, such as α-quartz,[9] α-HgS[10] and trigonal selenium (Se) and tellurium (Te),[11] have led to the discovery of novel phenomena that can be exploited in optoelectronic and spintronic devices[10,12–18] and to the development of new synthesis protocols to obtain these materials[2,15,19–21]. Among inorganic chiral materials, trigonal Te stands out due to its transport, optical and optoelectronic properties, which depend on the crystallographic orientation.[12,14] Indeed, it has been demonstrated that Te exhibits electrical magneto-chiral anisotropy and spin polarization,[16,22,23] electrical conductivity anisotropy and intrinsic polarized photoresponse,[23–25] tunable Rashba spin-orbit coupling,[26] optical activity[13,21,27] and bulk photovoltaic effect (BPVE)[18]. Given this rich phenomenology, recent efforts have focused not only on the development of Te nanostructures (*e.g.* nanowires, flakes),[12,15,16,19,28,29] but also on its growth with a preferential crystallographic orientation[13,30–32] or specific handedness[13,21]. In order to exploit the properties associated with crystal orientation and chirality in Te, the identification of the material's crystal faces and handedness is required. In this way, new research opportunities can be explored to develop a material-device platform capable of controlling charge, spin, and light. In this regard, the vibrational properties of Te [15,24,33–35] are not only an important feature linked to the chiral structure and lattice symmetry, but can facilitate the study of anisotropy, handedness and orientation of the crystal.

Trigonal tellurium presents enantiomer structures with right-handed (space group – $P3_121$, no. 152) and left-handed ($P3_221$, no. 154) helices formed by atomic chains arranged along the *c*-axis of the unit cell.[11,36] Preliminary studies of Te lattice dynamics by Raman spectroscopy were able to identify three active modes and determine their symmetry.[33,37–40] More recently, studies exploring the strong in-plane anisotropy of Te by linearly polarized Raman measurements were able to determine the arrangement of the helical chains in solution-grown 2D Te flakes and confirm the results by electron microscopy imaging.[15,24,41] Furthermore, by combining circularly-polarized Raman spectroscopy and first-principle calculations, the handedness of Te bulk crystals could be tentatively determined.[34,35] However, to exploit Raman spectroscopy to identify the crystal orientation and handedness, a study of different crystallographic planes is necessary, while in the literature only one plane is reported.[15,24,41] In this work, we investigate trigonal Te by polarized Raman spectroscopy and the corresponding symmetry rules in bulk crystals. Unlike Te flakes, wires and nanocrystals, bulk Te gives access to several crystallographic planes. Using XRD patterns, we are able to assign the different faces observed to specific crystal planes, correlating the polarized Raman results to the crystallographic orientation of the Te. We investigate three different crystallographic planes, one orthogonal to the chains and two parallel to them, by angle-dependent linearly polarized Raman spectroscopy and Raman tensor analysis. This complete dataset enables the measurement of the full Raman tensors for each mode, simplifying the identification of orientation from polarized Raman spectroscopy. Our results show that the angular dependence of the modes' intensity allows to distinguish the helical chains' arrangement and orientation. Furthermore, by circularly polarized Raman measurements, we univocally correlate the handedness of the crystals, verified by chemical etching, with the energy shift of the modes observed under opposite helicity configurations. The energy shift depends on the circular polarization configuration and can be robustly observed only for one incidence direction, highlighting the relevance of the proper crystal orientation for this measurement. This may explain why some recently reported optical phenomena such as BPVE are only detected when the illumination falls at a 45° angle to the helical chain plane in Te flakes[18] or why anisotropic chiroptical activity is observed in Te nanorods when changing their helical chain alignment with respect to the circular dichroism probe beam[13]. These optical phenomena are represented by their associated tensors,[42,43] and thus depend on the crystal symmetry and orientation. The latter can be assessed by polarized Raman spectroscopy.

**Experimental**

**Materials.** The Tellurium bulk crystals (5) used in this work were purchased from 2D Semiconductors and, as indicated by the supplier, they were synthesized through the flux zone growth technique with purity 99.9999% (6N).

**Chemical etching.** Tellurium bulk crystals were placed individually in glass Petri discs with 500 μL of $H_2SO_4$ (≥ 95%, Acros Organics) and heated for 30 min at 100°C on a hot plate (see **ESI**, **Fig. S1**). Afterwards, the crystals were washed with deionized water and dried with $N_2$ flow.

*Optical microscope images* of the crystals were taken with a Leica DM4000M microscope using different objectives from 5× to 50× in bright field.

**Raman spectroscopy characterization.** Micro-Raman characterization was carried out in a Renishaw® inVia micro-Raman instrument. For low-temperature measurements (220 K) and to minimize the air-exposure of the Te crystals, we used a 50× objective (N.A. 0.5) and a Linkam® chamber under $N_2$ flow equipped with a liquid $N_2$ cooling system. We collected the Raman spectra in the low-frequency range (edge filter at 30 cm$^{-1}$) using 532 nm as excitation source and a diffraction grating of 3000 l/mm. The laser power was kept below 0.1 mW to avoid damage to the crystals during the acquisitions,[44] which could be observed by the appearance of the $TeO_2$ related bands,[45] or heating-induced mode shift and broadening[46]. Angle-dependent linearly polarized measurements were carried out at room temperature rotating the samples in a piezo-motor (Thorlabs, ELL18/M rotation stage – PC controlled by ELLO® software) placed on the Raman instrument's XYZ stage and using a half-wave plate (Renishaw®) in the excitation path to rotate the polarization of the excitation laser, and a linear polarizer in the detection path. For the circularly polarized measurements, we combined a half-wave plate in front of the laser exit with a quarter-wave plate just before the Rayleigh filters in the excitation path (Renishaw®) and a quarter-wave plate (Thorlabs, AQWP10M-580) with a linear polarizer (Renishaw®) in the detection path. Python and Julia codes were employed to analyze the collected data, fitting the Raman spectra datasets to Lorentzian functions. We checked the reproducibility of our results by measuring 5 different crystals in several regions. We carried out control experiments to check the set-up and measurement conditions (**Figs. S2-S5**).

**X-ray diffraction characterization.** X-ray diffraction analysis was carried out on a Malvern-PANalytical 3$^{rd}$ generation Empyrean X-ray powder diffractometer. The instrument was equipped with a 1.8kW CuKα ceramic X-ray tube operating at 45 kV and 40 mA, iCore and dCore automated PreFIX optical modules, motorized Eulerian Cradle (χ-chi, ϕ-phi, x, y and z movements) and solid-state hybrid pixel PIXcel3D detector. The XRD patterns were collected in air at room temperature, using a zero-diffraction silicon substrate.

## Results and discussion

We studied five commercial bulk Te crystals (noted as C$i$ – C1 to C5) (see Experimental section), with trigonal crystal structure[11,36] confirmed by powder XRD (**Fig. 1**a-b). The XRD patterns from bulk crystals also allow us to assign the various faces observed to specific crystallographic planes. Insets in Fig. 1b show the macroscopic appearance of these faces: one mirror-like face and two matte faces. The XRD pattern of the mirror face (green in the scheme of Fig. 1b) shows peaks at 23, 47 and 73.5 degrees, corresponding to the *(h 0 0)* reflections, indicating that the crystallographic plane is (1 0 0). A negligible contribution from the (1 2 0) plane is also detected. In the case of the lateral face of the crystal (blue), reflections corresponding to the (1 1 0) and (2 2 0) planes at 40.5 and 87.4 degrees are observed. As in the mirror face, a reflection from the (1 2 0) is observed, which is more marked in this case, though still faint. This is due to the small angle difference between the (1 1 0) and the (1 2 0) planes (see **Fig. S6**). When measuring the last plane (red), the XRD pattern displays a single reflection peak, corresponding to the (0 0 3) plane. We thus observe three different crystallographic planes in bulk Te: two containing the *c*-axis and running parallel to the Te chains, planes (1 0 0) and (1 1 0), and another plane that is perpendicular to the chains and the *c*-axis, (0 0 3). Moreover, as mentioned in the introduction, trigonal Te presents enantiomer structures with right-handed (space group - $P3_121$ – no. 152) and left-handed ($P3_221$ – no. 154) helices formed by atomic chains arranged parallel to one another along the *c*-axis of a prismatic unit cell[11,36]. We determined the handedness of the five crystals under study by chemical etching[47]. After exposure to sulfuric acid, asymmetric etch pits appear on the (1 0 0) plane (mirror face) (Fig. 1c). The shape of the etch pits marks not only the chirality but also the direction of the helical chains (*c*-axis), which is useful for the subsequent spectroscopic characterization of the crystals. In each crystal (see **Fig. S1**), all the etch pits are aligned parallel to the *c*-axis (longer side) and point in the same direction, indicating same handedness in the whole plane. When moving from the *c*-axis to the opposite side, the rotation is different among the crystals. Crystals C3 and C4 rotate clockwise, thus, they are right-handed (RH); C1, C2 and C5 rotate anticlockwise, *i.e.*, they are left-handed (LH).

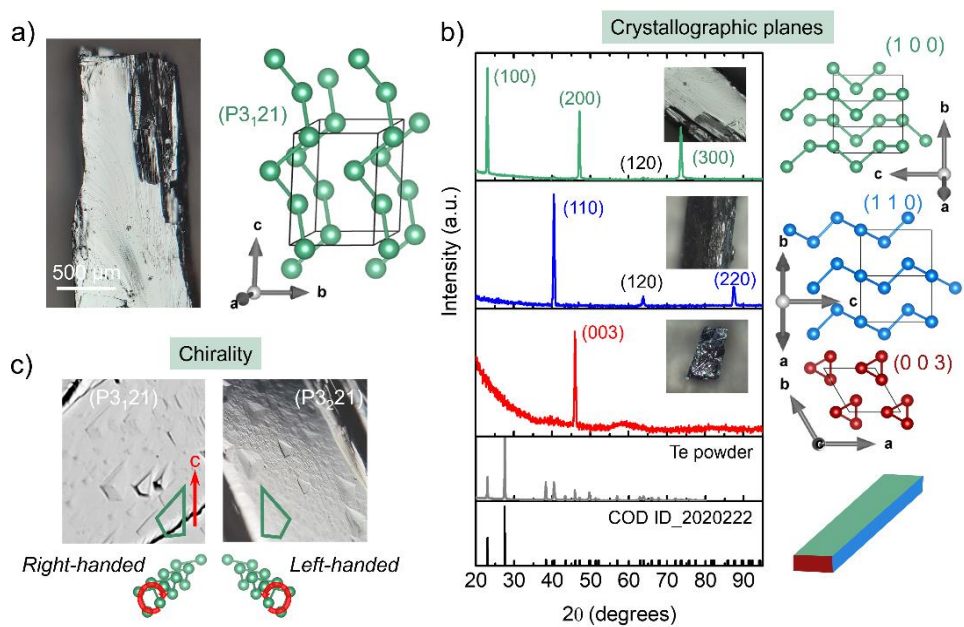

**Fig. 1.** a) Photograph of a representative bulk Te crystal showing the mirror-like face - crystallographic plane (1 0 0) - together with a sketch drawn using the crystallographic data of Adenis *et al.*[36] for a right-handed structure (space group – $P3_121$) with Vesta software. b) XRD patterns collected in a bulk crystal from the different accessible faces (green, blue and red color-code in the prismatic scheme, and photographs) and in powder form allowing to determine the crystallographic planes shown on the right. c) Optical images of two Te crystals of different handedness showing the etch pits produced by sulfuric acid treatment of the (1 0 0) plane and used to determine the chirality of the sample. The green polygon marks the outline of an etch pit, with its long side parallel to the *c*-axis.

The crystallographic orientation plays a role in the selection rules for Raman spectroscopy under linearly polarized light. Therefore, having identified the crystallographic planes, we carried out Raman spectroscopy measurements in the three crystal faces (**Fig. 2**). The Raman spectra collected from all crystal faces show three phonon modes corresponding to one $A_1$ singlet and two E doublets ($E^1$ and $E^2$), which are active modes according to the selection rules for the Te crystal's symmetry (point group $D_3$ – 32, Wyckoff position $3a$ (x, 0, 1/3 ó 2/3))[33,48–50]. The $A_1$ mode observed at ~120 cm$^{-1}$ corresponds to an expansion of the helical chain in the direction perpendicular to the *c*-axis in which each Te atom moves in the basal plane[39]. The E modes centered at ~92 and ~140 cm$^{-1}$ are ascribed to bending and stretching Te-Te bond vibrations[39]. To gain deeper insight into the symmetries of the three Raman modes observed and the directionality of Te's lattice vibrations, we performed angle-dependent linearly polarized Raman spectroscopy. We rotated the sample by an azimuthal angle θ, while keeping the polarization analyzer at the spectrometer and the polarization of the incident laser fixed. In this way, we could correlate the measured intensity dependence on θ with the crystal orientation. We carried out Raman spectroscopy measurements with incidence on the three faces of the crystal in parallel and cross polarization configurations: in the former the linear polarizers in excitation and detection paths were parallel (VV – **Fig. 2**a), whereas in the latter they were perpendicular (HV – **Fig. S7**a). Focusing on the crystallographic planes (1 0 0) and (1 1 0), where the helical chains are arranged parallel to the plane, we observed that the relative intensities of the three modes are sensitive to the sample orientation, and both planes present similar polar plots. As shown in the VV (Fig. 2c) and HV (Fig. S7) polar plots, the $A_1$ and E modes with a clear angular dependence highlight the strong in-plane anisotropy of the trigonal Te crystal. The polar plots provide a way to determine the direction of the helical chains (*c*-axis), confirmed by XRD and etching reported above. Specifically, in the parallel configuration (VV) the direction of the helical chains (*c*-axis) corresponds to the minima in the intensity of the three phonon modes, which occur on the diagonal 90°-270°. By looking into the lattice dynamics, this trend agrees with the symmetry of the modes, since in the case of the $A_1$ mode the chain expansion vibration, as well as most of the stretching vibrations of the E modes, occur in the *x*-direction (perpendicular direction to the *c*-axis). Indeed, researchers have observed this trend previously in the VV configuration for Te flakes in the (1 0 0) plane,[15,24,41] but not for other crystallographic planes as here we demonstrate. Focusing on the (0 0 3) plane (red plots)-hereafter noted as (0 0 1) since it belongs to the same family of planes (0 0 $\ell$), *i.e.*, with incidence along the *c*-axis (parallel to the helix axis)-, the shape of the polar plots for the phonon modes are different from the ones observed along the helical chains. In this case, the VV polar plots are isotropic for the $E^1$ and $A_1$ modes, and nearly isotropic for the $E^2$ mode. This reflects the anisotropy of the Te crystal structure when changing from the (1 0 0)/(1 1 0) to the (0 0 1) plane. However, it is not possible to extract information about the orientation of the Te-Te-Te triangles, since in this (0 0 1) plane the vibrations related to the three phonon modes are isotropic. Therefore, the angle-dependent linearly polarized Raman spectroscopy measurements shown here can be used as fingerprints to distinguish between the crystallographic planes parallel [(100)/(110) – equivalent planes] and perpendicular [(001)] to the helical chains, as well as to determine the orientation of the *c*-axis. To correlate the angular dependence of the different modes with the symmetry rules, we analyzed them using the Raman tensor theory. We observed a good match with our experimental data and highlight the contribution to the different crystallographic planes (see ESI – section Raman tensor analysis, **Figs. S8-S12**, **Table S1**). Our method to determine the Raman tensor can be extended to any crystal with point group $D_3$, $D_{3d}$ or $C_{3v}$.

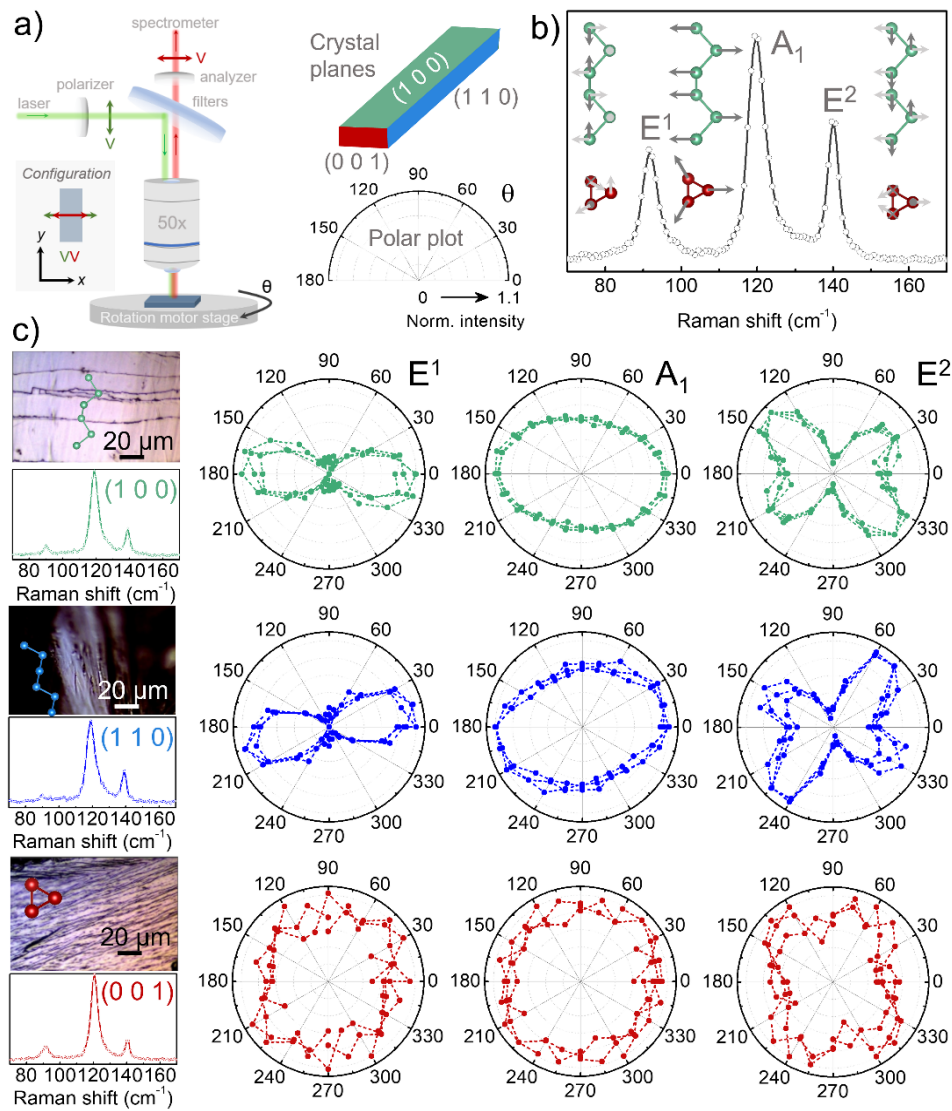

**Fig. 2.** Angle dependence of the phonon modes of the bulk Te crystals collected in parallel (VV) configuration: a) Sketch of the angle-dependent linearly polarized Raman measurements showing the optical elements for this configuration. Scale of the polar plots of the data obtained by rotating the Te bulk crystals (θ) and collecting the signal for each face (color-code) under study. b) Representative Raman spectrum of a bulk Te crystal (C2) showing the characteristic Raman modes ($E^1$, $A_1$ and $E^2$) and their corresponding atomic vibrations, viewed from the side and from above the chiral chains, as indicated by Martin et al.[39]. c) Optical images (above) taken with the micro-Raman instrument of the crystal planes under study indicating the arrangement of Te atoms and accompanied by the corresponding integrated Raman spectra along the polarization angle (below). Normalized polar plots of the three Raman modes' intensities (dots) extracted from the corresponding Raman spectra in at least three different regions of the images (in crystals C1 & C3). The orientation of the polar plots and optical images are the same, no correction to the rotation angle (θ) has been made.

Going a step further, the chiral symmetry of Te can be analyzed in detail using circularly polarized Raman spectroscopy. As a starting point, we carried out Raman spectroscopy with circularly polarized incident light while collecting the signal without any polarization optics, for the three crystal faces (**Fig. 3**a). When the incident and backscattered light is parallel to the *c*-axis of the Te crystals, *i.e.* falls upon the (0 0 1) plane, we observe a shift of the two E modes under opposite circularly polarized incident light ($\Delta E=E(L)-E(R)$, left – L and right – R), while the $A_1$ mode remains constant in frequency (Fig. 3b). Interestingly, the shift observed in the two E modes shows opposite sign depending on the handedness of the crystals. The RH crystals present a negative shift ($\Delta E^1 \sim -0.5$ cm$^{-1}$, $\Delta E^2 \sim -0.1$ cm$^{-1}$), while the LH crystals show a positive shift ($\Delta E^1 \sim +0.6$ cm$^{-1}$, $\Delta E^2 \sim +0.2$ cm$^{-1}$) for both the $E^1$ and $E^2$ modes. We found these results to be reproducible in different crystals (Fig. 3c and **Table S2**). It is important to note that the shift of the $E^1$ mode is considerably larger than that of $E^2$ mode, and can therefore be used more reliably to determine the handedness. Furthermore, when we carried out the same Raman characterization with incidence on the (1 0 0) and (1 1 0) planes (**Figs. S13** and **S14**, respectively), we did not observe any changes in the modes (**Tables S3-S4**). The shift observed in the E modes is in agreement with a previous study using the same polarization configuration (**Table 1**)[33]. In the present work, nonetheless, the handedness of the crystal was verified experimentally, supporting the result. This shift, induced by opposite circularly polarized incident light, results from the nature of E modes in the screw symmetry of the lattice,[33] where circularly polarized light interacts with the two components of the doublet in a different way, as discussed in detail in the following. In this work, we experimentally and univocally demonstrate the correlation between the sign of the observed shift and the handedness of the Te crystals. Moreover, we show that this shift is allowed, and therefore, observable with incidence along the *c*-axis - (0 0 1) plane.

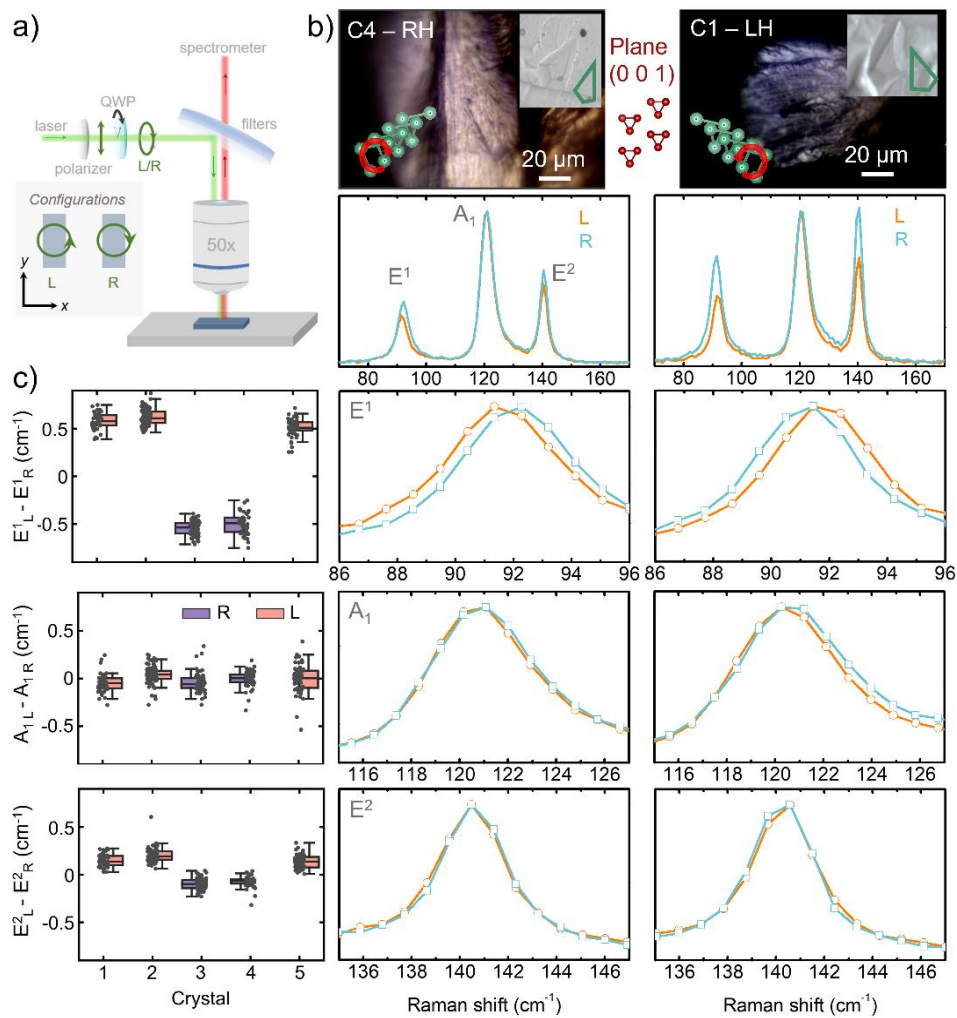

**Fig. 3.** Circularly polarized Raman characterization of representative right- and left-handed bulk Te crystals carried out falling upon the (0 0 1) crystallographic plane: a) Sketch of the circularly polarized Raman measurements showing the optical elements for this configuration using incident circularly polarized light and collecting unpolarized signal. b) Optical images taken with the micro-Raman instrument of the crystal faces under study accompanied by representative Raman spectra showing the characteristic Raman modes ($E^1$, $A_1$ and $E^2$) obtained using right- (R, blue) and left- (L, orange) handed light excitation (below). The insets show images of the corresponding etch pit indicating that the Te crystals handedness: right (C4) and left (C1). The panels below present zoomed ranges of the normalized Raman spectra making more visible the E phonon doublet shift while $A_1$ phonon does not change. c) Box and scatter plots (dots display the data collected) for the phonons shift under right- and left-handed light excitation for the five crystals under study: left- (C1, C2, C5) (pink) and right-handed (C3, C4) (purple).

Besides polarizing the incident light, we also checked helicity-resolved Raman spectroscopy, using opposite circularly polarized incident and collected light (LR vs RL, cross-helicity). Therefore, in this case, we added polarization optics in the collection path. We carried out the corresponding measurements on the three crystallographic planes (**Fig. 4**a). This configuration selects the angular momentum exchanged between photons and phonons, and thus detect "helicity-changing" Raman active modes, also known as "chiral" phonons. Experimentally, this can change the relative intensity of the peaks under different circular polarization configurations and results in a shift when incident and collected light have opposite circular polarization (i.e., the RL and LR cases). The underlying mechanism is related to the angular and pseudo-angular momentum (PAM) of the phonon, $m_{ph}$. Discrete rotational symmetry gives rise to conservation of PAM. [51–53] The incoming light can exchange angular momentum with the lattice vibrations, and in the case of discrete rotational symmetry (for Te, the $D_3$ point group with a 3-fold screw axis), the angular momentum can also be provided by the rotational periodicity of the lattice in a "rotational Umklapp" process.[52] The helicity change is thus allowed for degenerate E modes, but not for the non-degenerate $A_1$ mode. As a consequence of the conservation of PAM, in the RL and LR configuration the Raman scattering selects phonons with $m_{ph}$=+1 or -1. In tellurium, the degeneracy of the E modes is lifted in the phonon branches away from the center (Γ) of the Brillouin zone. In particular, in LH-Te the phonon with $m_{ph}$=-1 has lower energy than the one with $m_{ph}$=+1, and the opposite is true for RH-Te.[35] Furthermore, Raman scattering occurs at a small but finite momentum. Therefore, the E mode will appear at different energy under the LR and RL configurations, according to the $m_{ph}$ to fullfill the conservation rule. In particular, for LH-Te we expect the peak in RL to be at lower energy than LR, and the opposite for RH-Te. On the other hand, the non-degenerate $A_1$ mode ($m_{ph}$=0) is not affected. Indeed, when incident and scattered light are parallel to the c-axis of the Te crystals, *i.e.* falling upon the (0 0 1) plane, we observe a shift of the two E modes ($E^1$ and $E^2$), while $A_1$ mode remains at the same frequency. Moreover, the shift observed in both $E^1$ and $E^2$ modes (defined as $\Delta E = E(LR)-E(RL)$) shows opposite sign depending on the handedness of the crystals: RH crystals present a negative sign ($\Delta E^1 \sim -0.8$ cm$^{-1}$, $\Delta E^2 \sim -0.2$ cm$^{-1}$), while the LH crystals show a positive sign ($\Delta E^1 \sim +0.9$ cm$^{-1}$, $\Delta E^2 \sim +0.2$ cm$^{-1}$). These results were found to be reproducible in sign and magnitude in different crystals (Fig. 4b-c and Table S2). Therefore, the E phonons behave as "helicity-changing" Raman modes, while the $A_1$ phonon is a "helicity-conserving" mode. Interestingly, the shift in the cross-helicity condition is larger than in the case of circularly polarized Raman shown before. In the latter configuration we observe a narrower signal (e.g., for mode $E^1$, full width half maximum, FWHM is ~5 cm$^{-1}$ vs 4.2 cm$^{-1}$ in cross-helicity), which could favor the resolution of the shift. Nonetheless, the set-up using circularly polarized light only in the incident beam (Fig. 3a) is sufficient for a definite assessment of the chirality of the crystal. We also note that this difference can explain the different values reported in the literature (Table 1). Furthermore, we checked that this behavior does not apply when incident and scattered light fall upon the (1 0 0) and (1 1 0) planes (**Figs. S15-S16** and Tables S3-S4, respectively).

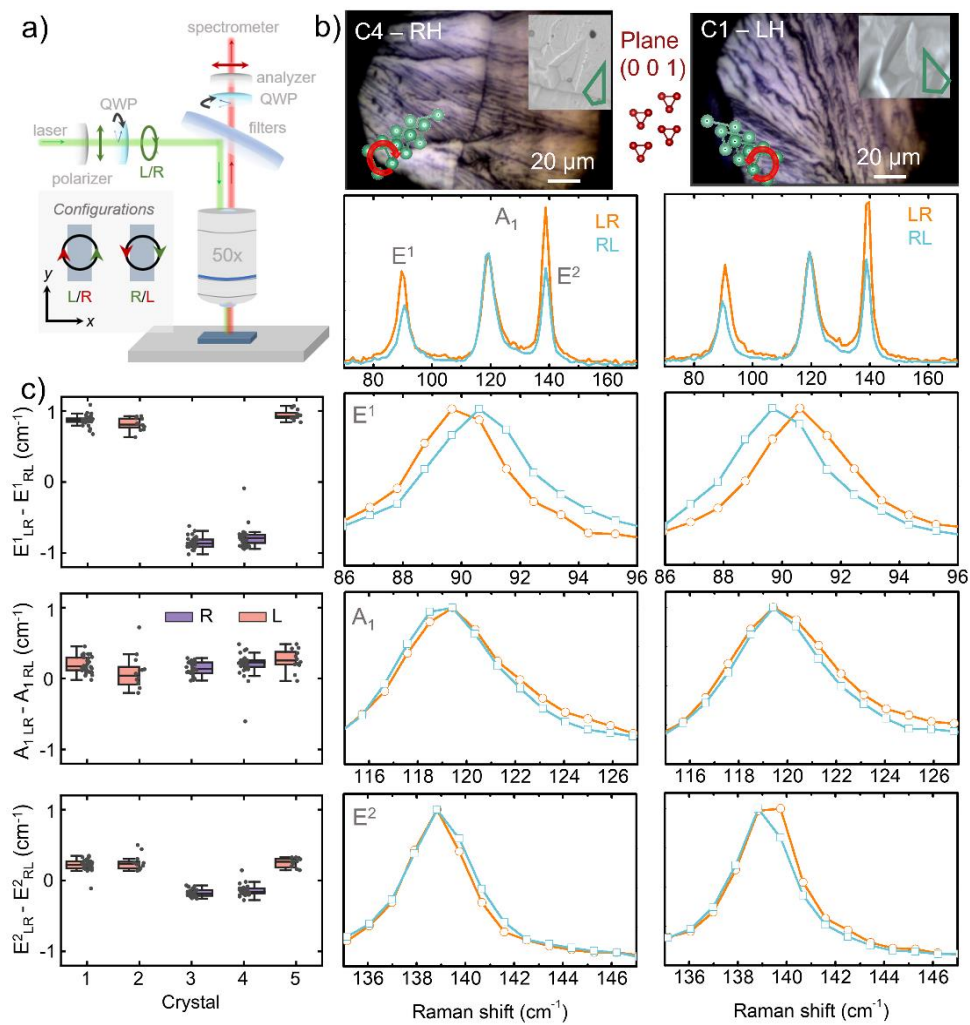

**Fig. 4.** Helicity-resolved Raman spectroscopy characterization of representative right- and left-handed bulk Te crystals carried out on the (0 0 1) crystallographic plane: a) Sketch of the Raman set-up showing the optical elements for this configuration using incident and collected circularly polarized light with opposite handedness: RL & LR. b) Optical images taken with the micro-Raman instrument of the crystal faces under study accompanied by representative helicity-resolved Raman spectra showing the characteristic Raman modes ($E^1$, $A_1$ and $E^2$) obtained using the two configurations: RL & LR. The insets show images of the corresponding etch pit indicating the Te crystals' handedness: right (C4) and left (C1). The panels below present zoomed ranges of the normalized Raman spectra to better illustrate how the E phonon doublet shifts while $A_1$ does not change. c) Box plots accompanied by the experimental data (dots) showing the phonons shift under RL and LR configurations for the five crystals under study: left- (C1, C2, C5) and right-handed (C3, C4).

Additionally, we note that the Raman tensor formalism, which was applied to the analysis of the linearly polarized spectra, can also be suitable for the circularly polarized spectra. To consider states of opposite PAM at finite wavenumber, the scattering intensity from E(x) and E(y) (see ESI) must not be added together. In the case of linear polarization, both modes are excited, so it is appropriate to add the two terms. The tensor coefficients *c,d* for the two modes deriving from the degenerate E mode will not, in general, be the same. However, since the scattering occurs for momentum very close to Γ, we expect that the coefficients will be very similar for the two cases. In particular, we expect the scattering for the E modes in both LR and RL configuration to be proportional to $c^2$. The case of the experiment with only incident circular polarized light follows the same trend and can be calculated choosing the suitable polarization in the Raman tensor formalism.

**Table 1.** Comparison with the literature of the shift of the E modes (absolute value) when using different circularly polarized Raman configurations in bulk Te crystals with light incident on the (0 0 1) plane.

| References | $E^1$ mode shift (cm$^{-1}$) | $E^2$ mode shift (cm$^{-1}$) | Configuration | Laser source (nm) |
|---|---|---|---|---|
| Pine & Dresselhaus[33] | 0.6 | 0.3 | R/- & L/- | 514.5 |
| Ishito et al. [35] | 0.7-0.8 | 0.2-0.3 | RL & LR (PAM) | 785 |
| Zhang et al. [34] | 1.15-1.08 | 0.38-0.28 | RL & LR (PAM) | 633 |
| This work | 0.56±0.05 | 0.13±0.05 | R/- & L/- | 532 |
|  | 0.85±0.06 | 0.21±0.04 | RL & LR (PAM) | 532 |

## Conclusions

We presented a comprehensive and detailed study of trigonal Te bulk crystals by linear and circularly polarized Raman spectroscopy, showing that this technique allows experimentally identifying the orientation and handedness of this material. We demonstrated that angle-dependent linearly polarized Raman spectroscopy in parallel configuration allows determining the direction of the helical chains (*c*-axis) in the (1 0 0) and (1 1 0) planes, reflected as minima in the intensity angular dependence of the three active modes, $E^1$, $A_1$ and $E^2$. Moreover, comparing the results obtained for the three Te faces examined, we showed that it is possible to distinguish the crystallographic planes that are parallel to the helical chains, i.e. (1 0 0) and (1 1 0), from those that are perpendicular, i.e. (0 0 1), by observing the angular dependence of the Raman mode intensity, which is anisotropic in the former case and isotropic in the latter. The angular dependence and alignment of the crystal was supported by Raman tensor analysis. Furthermore, by circularly polarized Raman measurements, we correlated the handedness of Te crystals, determined by chemical etching, with the sign of the shift observed in the $E^1$ and $E^2$ phonons. More specifically, we demonstrated and confirmed in several crystals that: (i) the magnitude of the shift depends on the circularly polarized Raman configuration used, giving consensus to the different values reported in literature and proving that just circularly polarized light in the incident light is enough to observe the phonon shift; (ii) the shift of the $E^1$ mode is larger than that of $E^2$, and is therefore considered more sensitive and reliable when trying to determine the crystal handedness; and (iii) only by incidence parallel to the helical chains, plane (0 0 1), is it possible to determine the handedness of the Te crystal; the other planes (1 0 0) and (1 1 0) cannot be used for this purpose.

The findings and approach presented in this work can be applied to other anisotropic and chiral materials to gain insight into the crystal structure and symmetry using Raman spectroscopy and a detailed polarization analysis. Moreover, we highlight that the response under linearly and circularly polarized light depends on the direction of incidence with respect to different crystal faces, which has implications when selecting the proper crystal orientation for optical and electrical transport studies. In this way, a guideline for the correlation of device design with crystal growth protocols is supported by Raman spectroscopy.

## Author Contributions

Davide Spirito: Conceptualization, Investigation: Raman spectroscopy measurements, Raman tensor analysis, Resources, Data curation, Validation, Formal analysis, Visualization, Writing - Reviewing and Editing. Sergio Marras: Investigation: XRD characterization, Resources, Data curation, Writing - Reviewing and Editing. Beatriz Martín-García: Conceptualization, Supervision, Methodology, Investigation: Raman spectroscopy measurements, Resources, Data curation, Validation, Formal analysis, Visualization, Writing - original draft preparation, Writing - Reviewing and Editing, Project administration, Funding acquisition.

**Conflicts of interest**

There are no conflicts to declare.

**Electronic Supplementary Information (ESI)**

Photograph of a Te crystal during the chemical etching process and optical images of the etch pits for the Te crystals understudy; Control experiments regarding Raman data acquisition conditions and polarization tests in silicon (100); Angle-dependent linearly polarized Raman spectroscopy measurements in cross configuration; Raman tensor analysis for linearly polarized Raman spectroscopy; and Circularly polarized Raman spectroscopy measurements in the (100) and (110) planes. See https://doi.org/10.1039/D3TC04333A

**Acknowledgements**

This work is supported under Project PID2021-128004NB-C21 funded by Spanish MCIN/AEI/10.13039/501100011033 and under the María de Maeztu Units of Excellence Programme (Grant CEX2020-001038-M). Additionally, this work was carried out with support from the Basque Science Foundation for Science (IKERBASQUE). B.M.-G. thanks support from "Ramón y Cajal" Programme by the Spanish MCIN/AEI (grant no. RYC2021-034836-I) and IKERBASQUE HYMNOS project. B.M-G. thanks to Prof. A. Mateo-Alonso (Molecular and Supramolecular Materials Group - POLYMAT) for the access to the Chemistry Lab to carry out the chemical etching of the crystals. Authors thank Prof. L. E. Hueso and Dr. E. Goiri Little (Nanodevices group – CIC nanoGUNE) for reading and revising the manuscript.